\documentclass[sigconf]{acmart}

\AtBeginDocument{%
  }

\setcopyright{acmlicensed}

\copyrightyear{2025}
\acmYear{2025}
\setcopyright{cc}
\setcctype{by}
\acmConference[DeFi '25]{Proceedings of the 2025 Workshop on Decentralized Finance}{October 13--17, 2025}{Taipei, Taiwan}
\acmBooktitle{Proceedings of the 2025 Workshop on Decentralized Finance (DeFi '25), October 13--17, 2025, Taipei, Taiwan}\acmDOI{10.1145/3733815.3764043}

\acmISBN{979-8-4007-1904-2/2025/10}

\usepackage{pifont}
\usepackage{colortbl}
\usepackage{xcolor}
\usepackage{enumitem}
\usepackage{colortbl}
\usepackage{multirow}
\usepackage{makecell}
\usepackage{subcaption}

\definecolor{lightgray}{gray}{0.95}

\begin{document}

\title[Cross-Chain Sealed-Bid Auctions]{Cross-Chain Sealed-Bid Auctions Using Confidential Compute Blockchains}

\author{Jonas Gebele}
\affiliation{%
  \institution{Technical University of Munich}
  \city{Munich}
  \country{Germany}}
\email{jonas.gebele@tum.de}

\author{Timm Mutzel}
\affiliation{%
  \institution{Technical University of Munich}
  \city{Munich}
  \country{Germany}}
\email{timm.mutzel@tum.de}

\author{Burak Öz}
\affiliation{%
  \institution{Technical University of Munich}
  \city{Munich}
  \country{Germany}}
\email{burak.oez@tum.de}

\author{Florian Matthes}
\affiliation{%
  \institution{Technical University of Munich}
  \city{Munich}
  \country{Germany}}
\email{matthes@tum.de}

\renewcommand{\shortauthors}{Gebele et al.}

\begin{abstract}
Sealed-bid auctions ensure fair competition and efficient allocation but are often deployed on centralized infrastructure, enabling opaque manipulation. Public blockchains eliminate central control, yet their inherent transparency conflicts with the confidentiality required for sealed bidding. Prior attempts struggle to reconcile privacy, verifiability, and scalability without relying on trusted intermediaries, multi-round protocols, or expensive cryptography. We present a sealed-bid auction protocol that executes sensitive bidding logic on a Trusted Execution Environment (TEE)-backed confidential compute blockchain while retaining settlement and enforcement on a public chain. Bidders commit funds to enclave-generated escrow addresses, ensuring confidentiality and binding commitments. After the deadline, any party can trigger resolution: the confidential blockchain determines the winner through verifiable off-chain computation and issues signed settlement transactions for execution on the public chain. Our design provides security, privacy, and scalability without trusted third parties or protocol modifications. We implement it on SUAVE with Ethereum settlement, evaluate its scalability and trust assumptions, and demonstrate deployment with minimal integration on existing infrastructure.
\end{abstract}




\maketitle

\section{Introduction}

Auctions are core mechanisms for incentive-compatible resource allocation and price discovery, particularly in markets with asymmetric information or changing valuations. In first-price sealed-bid auctions, participants privately submit bids, and the highest bidder wins, paying the amount they bid. This auction format is deployed in high-stakes settings such as spectrum licensing~\cite{Cramton2002, Ihle2018}, carbon credit allocation~\cite{McAdams2011}, and real-time digital advertising~\cite{Goke2024}.
In practice, auction systems often rely on centralized operators with unilateral control over bidding logic, data access, and settlement, enabling opaque deviations from sealed-bid semantics that are hard to detect. For example, \textnormal{Google}’s advertising infrastructure, which generated over \$260 billion in revenue in 2023~\cite{Bianchi2025}, has been at the center of multiple antitrust investigations. From 2013 to 2019, the platform retained surplus value from second-price auctions by withholding information about the actual second-highest bid and reallocating that surplus to favor its own ad-buying tools, thereby distorting competition~\cite{2021}. In 2018, a secret arrangement with \textnormal{Facebook} granted preferential access to bid information, reduced transaction costs, and guaranteed winning rates, despite claims that the system operated as a neutral, sealed-bid auction~\cite{Chee2022}. Across these cases, centralized control permitted outcome manipulation that remained largely invisible to participants and regulators.

Public blockchains provide a compelling infrastructure for decentralized auctions by enabling verifiable execution of auction rules and non-custodial asset transfers  without relying on trusted intermediaries. However, this transparency inherently conflicts with the confidentiality requirements of sealed-bid auctions. On-chain interactions such as registering for the auction or submitting bids expose bid values, timing, and funding patterns, which can be linked to pseudonymous identities and exploited for front-running or strategic inference. Additionally, the visibility of bidder activity near auction deadlines exposes them to targeted censorship by adversarial validators.

\begin{figure*}[t!]
    \centering
    \begin{subfigure}[t]{0.49\textwidth}
        \centering
        \includegraphics[width=\linewidth]{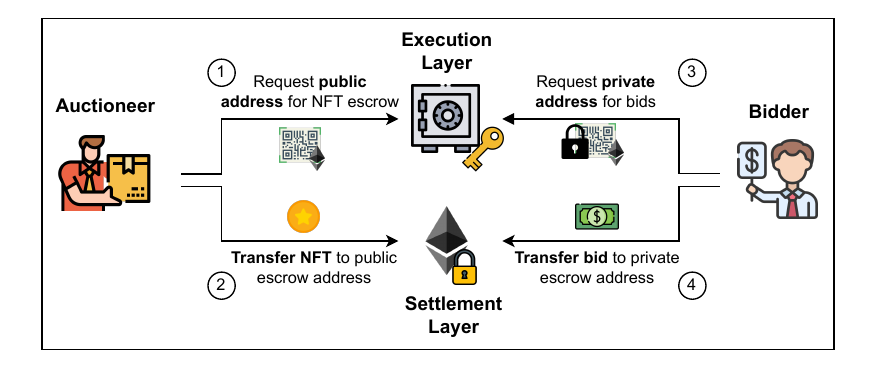}
        \caption{Auction Setup and Commitment.}
        \label{fig:auction-setup}
    \end{subfigure}
    \hfill
    \begin{subfigure}[t]{0.49\textwidth}
        \centering
        \includegraphics[width=\linewidth]{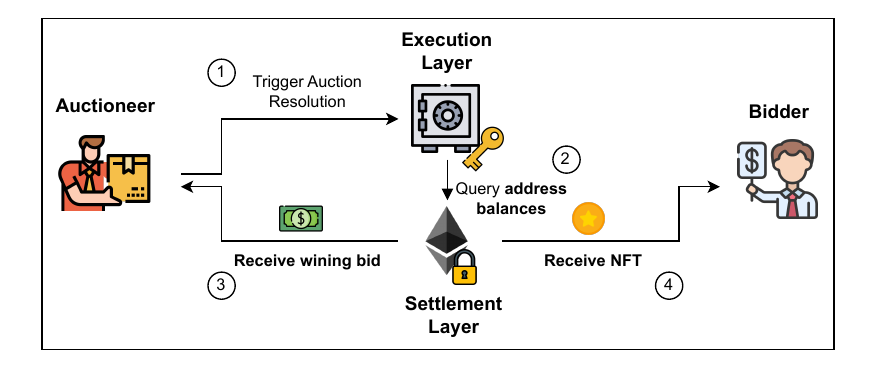}
        \caption{Confidential Winner Selection.}
        \label{fig:winner-selection}
    \end{subfigure}
    \caption{Protocol flow of confidential sealed-bid auctions.  
    (\subref{fig:auction-setup}) The auctioneer escrows the asset to a public blockchain address controlled by the confidential-compute blockchain; bidders receive encrypted, per-bidder funding addresses for private commitment.  
    (\subref{fig:winner-selection}) After the bidding deadline, the confidential compute blockchain verifies escrow balances via off-chain queries, determines the highest valid bid, and emits signed settlement transactions for on-chain execution.}
    \label{fig:auction-intro}
\end{figure*}

Various approaches have been proposed to realize sealed-bid auctions on blockchains, spanning cryptographic constructions, protocol-level designs, and hardware-based techniques. Yet all known designs face fundamental tensions between privacy, decentralization, and enforceability. Ensuring bid confidentiality often involves costly zero-knowledge proofs~\cite{Galal2018}, coordination-heavy protocols~\cite{Blass2017}, or dependence on centralized trusted hardware~\cite{Galal2019}. Conversely, designs that avoid such overheads tend to rely on centralized actors~\cite{Kosba2016}, introduce complex interactive procedures~\cite{Tyagi2023}, or weaken censorship resistance ~\cite{Tyagi2023}. Despite significant progress, existing systems still fall short of achieving sealed-bid privacy without introducing central points of failure or imposing prohibitive technical burdens~\cite{Zhang2024}.

To address these limitations, we leverage confidential compute blockchains that embed Trusted Execution Environments (TEEs) at the execution layer. \emph{Confidential compute} blockchains typically decouple private execution from consensus by offloading sensitive contract logic to hardware-isolated TEEs. Encrypted user inputs are processed inside these enclaves, which maintain an confidential persistent state and produce signed, attestable outputs. The consensus layer handles ordering and data availability, enabling privacy-preserving smart contract execution without revealing sensitive information on-chain. Notable implementations include Secret Network~\cite{SNT2020}, Oasis Sapphire~\cite{OasisFoundation2020}, Phala Network~\cite{Yin2022}, TEN Network (formerly Obscuro)~\cite{TenFoundation2023}, and Flashbots’ SUAVE~\cite{FRET2023}.

Our protocol preserves bid confidentiality by offloading privacy-sensitive logic to a confidential compute execution chain, while delegating asset custody and final settlement to a public blockchain such as Ethereum. The auction logic like bid registration and winner determination is executed off-chain within secure enclaves, and the resulting outcome enforced on-chain. To ensure enforceability, the execution chain takes control over the committed assets by provisioning per-bidder escrow addresses on the settlement layer.

In the protocol illustrated in Figure~\ref{fig:auction-intro}, the auctioneer first transfers the auctioned asset to a public escrow address generated by the execution chain (\subref{fig:auction-setup}). Bidders then request encrypted, per-bidder escrow addresses which they fund on the settlement layer to privately commit their bids. After the auction deadline, the execution chain determines the winning bid by verifying bidders’ settlement-layer balances through verifiable off-chain queries executed inside the enclave (\subref{fig:winner-selection}). After resolution, the execution chain emits signed settlement transactions, which reveal the set of participating addresses and corresponding bid transfers. This enables public auditability without compromising bid confidentiality prior to resolution.

The protocol provides strong guarantees without relying on a trusted third party. Interaction is limited to address registration and bid submission, eliminating multi-phase coordination. Bids are \emph{sealed} and \emph{binding}, as each commitment is bound to a bidder-specific escrow address managed by the enclave. \emph{Bidder privacy} is preserved until resolution, while outcomes are \emph{publicly verifiable} via attestable execution and \emph{auditable} through the published settlement transactions. Security rests solely on the integrity of the TEE and the execution guarantees of the confidential compute chain.

\vspace{0.5em}
\noindent\textbf{Contributions.}
This work makes the following contributions:
\begin{itemize}
\setlength\itemsep{0.6em}
\renewcommand\labelitemi{--}
\item We present a sealed-bid auction protocol that offloads sensitive auction logic to a confidential compute blockchain while performing settlement on a public blockchain, enabling private bidding, binding commitments, and trustless settlement with minimal bidder interaction.
\item We implement the protocol\footnote{\url{https://github.com/TimmMu/tee-confidential-auctions}} using existing infrastructure and demonstrate its deployability with minimal integration effort and on-chain overhead.
\item We evaluate the protocol’s security under standard assumptions on TEEs and confidential compute blockchains, showing that it satisfies core properties such as confidentiality, enforceability, and verifiability.
\item We analyze the trust and scalability tradeoffs, demonstrating that reliance on trusted hardware remains minimal, auditable, and does not reintroduce centralized control.
\end{itemize}

\medskip

\noindent\textbf{Paper Outline.}
Section~\ref{sec:background} introduces TEEs and describes how they are integrated into confidential compute blockchains for private off-chain computation. Section~\ref{sec:problem-statement} defines the system model, design goals, and adversarial assumptions underlying the protocol. Section~\ref{sec:protocol} outlines the protocol in detail, including its lifecycle and resolution logic. Section~\ref{sec:implementation} describes our prototype implementation, execution-layer requirements, and the protocol’s scalability characteristics. Section~\ref{sec:relatedwork} compares our protocol with existing approaches and evaluates how they align with the design goals. Finally, Section~\ref{sec:conclusion} discusses trade-offs and outlines paths toward reducing trust in confidential compute infrastructure.

\section{Background}
\label{sec:background}  

This section outlines the privacy limitations of on-chain smart contracts, introduces TEEs for secure off-chain computation, and explains how confidential compute blockchains integrate TEEs to enable verifiable, privacy-preserving execution.

\subsection{Blockchain \& Smart Contracts}

Blockchains are decentralized ledgers that maintain a consistent, tamper-evident transaction history across a peer-to-peer network. Transactions are grouped into blocks, each cryptographically linked to its predecessor, forming an append-only chain. Consensus protocols ensure all honest nodes agree on a single canonical chain.

Bitcoin~\cite{Nakamoto2008} introduced this model for decentralized digital currency, providing a secure ledger without central authority. However, it supports only limited scripting and fixed-functionality transactions. Ethereum~\cite{Wood2014} generalized blockchain programmability through the Ethereum Virtual Machine (EVM), which supports Turing-complete smart contracts with persistent state and composable logic. Ethereum uses an account-based model with two account types: externally owned accounts (EOAs), controlled by private keys, and contract accounts, governed by immutable on-chain code. Smart contracts reside at fixed addresses and expose callable functions that can be invoked by EOAs or other contracts. Their execution is deterministic and replicated across all validating nodes to ensure consensus.

This execution model enables trustless, verifiable enforcement of logic and asset transfers, but inherently exposes all operational details. Contract code, inputs, intermediate states, and outputs are publicly visible and deterministically executed across all validating nodes - ensuring consensus, but undermining confidentiality. In sealed-bid auctions, such transparency reveals bid amounts, identities, and timing patterns, enabling front-running, and strategic manipulation.

To overcome the confidentiality limitations of transparent execution, an emerging blockchain system architecture integrates hardware-based privacy into the execution layer. By leveraging TEEs, this model enables confidential smart contract execution over shared private state, while preserving verifiability and decentralization.

\subsection{Trusted Execution Environments}

TEEs enable confidential computation with integrity guarantees on untrusted hosts. 
Code executes inside an isolated enclave, a protected CPU context inaccessible to the operating system, hypervisor, or other processes. Data is automatically encrypted when transferred outside the enclave to prevent external access or modification. Intel Software Guard Extensions (SGX)~\cite{Anati2013, McKeen2013} is the most widely deployed TEE platform. It supports the execution of general-purpose code inside enclaves with hardware-enforced isolation. Data transferred outside the enclave is encrypted by a dedicated memory encryption engine. Enclaves cannot perform system calls directly but interact with untrusted components through controlled interfaces. Remote attestation allows an enclave to prove its identity and initial state to external parties using a hardware-signed cryptographic report. This mechanism enables secure input provisioning and verifiable output generation without relying on prior trust~\cite{Brickell2010}. SGX relies on Intel’s EPID signature scheme and proprietary verification infrastructure.
Although TEEs provide strong isolation, they do not ensure availability and have been subject to various attacks, including side-channel analysis, fault injection, and attestation forgery~\cite{Bulck2018, Moghimi2017, Chen2019, Schwarz2017, Cloosters2020, Mishra2020, Chen2023}. These limitations call for careful system design but do not preclude the use of TEEs as a practical foundation for confidentiality - provided the surrounding protocol can tolerate compromise without undermining integrity.

\subsection{TEE-based Confidential Compute Blockchains}

Several blockchain platforms have explored architectures for confidential smart contract execution. Traditional cryptographic techniques, such as zero-knowledge proofs, fully homomorphic encryption (FHE), and secure multiparty computation (MPC) - offer strong theoretical privacy guarantees, but suffer from practical limitations. FHE and MPC, in particular, entail high computational overhead and intricate setup procedures. While systems like Hawk~\cite{Kosba2016}, zkhawk~\cite{Banerjee2023}, Zkay~\cite{Steffen2020}, ZeeStar~\cite{Steffen2023} and Zapper~\cite{Steffen2022} demonstrate the feasibility of privacy-preserving contracts using these primitives, they face scalability and usability challenges that restrict their real-world deployment.

An emerging blockchain architecture for privacy-preserving smart contracts integrates TEEs into the execution layer to support private and verifiable off-chain computation. User inputs are encrypted using an attested enclave public key, allowing contract logic to operate over confidential state without exposing sensitive data on-chain. Execution within the TEE maintains data confidentiality and yields verifiable outputs via remote attestation, ensuring that results originate from an authenticated and integrity-protected program instance.

TEE-based confidential compute blockchains follow two main architectural models. One integrates TEEs directly into validator nodes, as in designs like Secret Network~\cite{SNT2020} and TEN Network~\cite{TenFoundation2023}. In this setting, each consensus node executes private contract logic within an enclave. This enables synchronous confidentiality and simplifies coordination, but increases execution overhead and enlarges the trusted computing base (TCB) to include all validators.
A second model decouples confidential execution from consensus by offloading private computation to TEE-equipped nodes. Results are attested and returned on-chain for verification. This architecture was first formalized in Ekiden~\cite{Cheng2019} and later improved in RaceTEE~\cite{Zhang2025} and forms the basis for systems such as SUAVE~\cite{FRET2023} and Oasis Protocol~\cite{OasisFoundation2020}. Related designs include the Phala Network~\cite{Yin2022}, and COMMIT-TEE~\cite{Erwig2023}, each of which explores different integration trade-offs. By separating execution from consensus, these systems reduce on-chain overhead and narrow the TCB.

To mitigate the impact of potential enclave compromise, many frameworks incorporate protocol-level safeguards such as key rotation, sealed-state re-encryption, and attestation freshness checks. These mechanisms offer forward secrecy and reduce long-term risk, ensuring that the compromise of a single enclave does not jeopardize system-wide confidentiality or integrity.


\section{Problem Setting and Design Goals}
\label{sec:problem-statement}  

We formalize the problem setting of realizing sealed-bid auctions on public blockchains with strong confidentiality and enforceability guarantees. Our goal is to design a protocol that preserves bid privacy, ensures verifiable resolution, and enforcable settlement, without relying on a trusted third party or modifying existing blockchain protocols.

\subsection{System Model}

The protocol operates across two decoupled blockchain layers and involves five distinct actors. These include the auctioneer, bidders, a confidential compute execution layer, a public blockchain settlement layer, and supporting settlement layer interfaces - summarized in Table~\ref{tab:actors}. For concreteness, we assume the auctioned asset is a non-fungible token (NFT) conforming to the ERC-721 standard~\cite{Entriken2018}.

\begin{table}[t]
  \centering
  \small
  \renewcommand{\arraystretch}{1.2}
  \setlength{\tabcolsep}{6pt}
  \begin{tabular}{@{}p{2cm}p{5.7cm}@{}}
    \toprule
    \textit{Actor} & \textit{Role in the Protocol} \\
    \midrule
    Auctioneer & Deploys the auction contract to the execution layer and escrows the NFT to a settlement-layer address controlled by the enclave. Has no further role or access. \\
    
    Bidders & Obtain a bidder-specific escrow address from the execution layer and fund it on the settlement layer. Identities remain hidden until resolution. \\
    
    Execution Layer & Confidential compute blockchain that executes auction logic on encrypted inputs, maintains sealed state, and emits verifiable outputs. \\
    
    Settlement Layer & Public blockchain that handles escrow, transaction ordering, and execution of enclave-signed settlement transactions. \\
    
    Settlement Layer Interface & Public interface endpoints to the settlement layer, enabling off-chain balance queries and submission of signed transactions from the execution layer. \\
    
    \bottomrule
  \end{tabular}
  \caption{Actors and their respective roles in the auction protocol architecture.}
  \label{tab:actors}
\end{table}

\paragraph{Data Flow and Communication Model.} To coordinate sealed bidding and settlement across chains, the protocol relies on a set of structured data flows:

\begin{itemize}[leftmargin=1.5em, itemsep=0.3em]
    \item[(i)] \textbf{Encrypted inputs} from bidders to the execution enclave, using an enclave-exposed public key to ensure that bid content remains confidential and is only accessible within the enclave.

    \item[(ii)] \textbf{Encrypted outputs} from the enclave to external recipients. By encrypting enclave-emitted data under recipient-specific keys, the protocol enables confidential, authenticated communication of bidder-specific messages in both directions.

    \item[(iii)] \textbf{Verifiable off-chain queries} from the enclave to public settlement-layer interfaces. These enable retrieval of external state without requiring native blockchain integration, and are supported by many confidential compute platforms via cross-chain messaging protocols (see Table~\ref{tab:exec-layer-capabilities}).
    
    \item[(iv)] \textbf{Public messages} emitted by the execution environment like signed settlement instructions or protocol results.
\end{itemize}

\subsection{Security Assumptions}
\label{sec:sec-assumptions}

This section details the trust model and security assumptions for each system element involved in the protocol.

\paragraph{TEE}
We assume a hardware root of trust with unforgeable attestation keys and remote-attestation reports that are verifiable. Under normal conditions, the TEE preserves code confidentiality and safeguards its internal state against tampering. Enclave memory is assumed to provide confidentiality, integrity, and rollback protection. Randomness is sourced from hardware instructions (e.g., \texttt{RDRAND}) and used for protocol key generation. We adopt a limited-leakage model in which side channels may reveal access patterns, but assume that full key extraction or integrity compromise remains rare in practice. Known attack vectors, including side channels, rollback exploits, and attestation subversion, are acknowledged (see Section ~\ref{sec:threats} and discussed (see Section~\ref{sec:conclusion}). A breach of a single enclave exposes settlement keys, breaking the auction and potentially stealing all funds.

\vspace{0.3em}
\paragraph{Execution Layer.}
All TEE-related assumptions extend to this layer. We assume a Byzantine fault-tolerant consensus that ensures availability, liveness, and consistent transaction ordering. Spam resistance and Sybil deterrence rely on gas costs, economic constraints, or validator-level filtering.

\vspace{0.3em}
\paragraph{Settlement Layer.}
We assume a standard public blockchain with eventual finality and liveness in the sense of Nakamoto-style consensus. While validators may delay, reorder, or censor transactions temporarily, finalized blocks are irreversible.

\vspace{0.3em}
\paragraph{Auctioneer and Bidders.}
All participants are assumed to behave arbitrarily. The auctioneer may abandon the auction, choose not to call resolution, or fail to relay settlement transactions. Bidders may collude, or spawn Sybil identities. Actors on both layers may observe, delay, reorder, or censor messages but cannot forge enclave outputs or access confidential data.

\vspace{0.3em}
\paragraph{Settlement Layer Interfaces.}
We assume that communication with interfaces occurs over authenticated channels. Some interfaces may behave maliciously by returning incorrect data or withholding responses. If multiple queried interfaces collude, correctness cannot be guaranteed. However, misbehavior is auditable, and further mitigation strategies are discussed in Section~\ref{sec:threats}.

\subsection{Protocol Goals and Guarantees}
\label{sec:goals}

We formalize the security, usability, and scalability properties that our sealed-bid auction protocol must satisfy trustlessly in decentralized settings.

\begin{enumerate}[leftmargin=2em, label=\textbf{\arabic*.}]
  \item \textit{Confidentiality.}  
  No adversary monitoring the execution or settlement layer should be able to infer the value, origin, or content of any valid bid prior to resolution.

  \item \textit{Binding.}  
  Each bid constitutes a verifiable commitment to transfer a specific value. Once submitted, it remains enforceable and cannot be revoked, overwritten, or redirected by the bidder.

  \item \textit{Non-Malleability.}  
  Each bid is bound to its originator and resistant to tampering. No party should be able to forge, modify, or reassign another bidder’s commitment.

  \item \textit{Enforceability.}  
  A valid auction must deterministically resolve, and its outcome must be executable - without reliance on any trusted party or centralized actor.

    \item \textit{Non-Interactivity.}  
    Each bidder participates through a single submission, without requiring multi-phase coordination or further interaction with other parties.
  
  \item \textit{Auditability and Verifiability.}  
  Any observer should be able to verify the correctness of the outcome using publicly available information and attestable execution results.

  \item \textit{Scalability.}  
    Per-bidder costs remain constant, independent of total auction size. Resolution incurs at most linear overhead with the number of bids.

\end{enumerate}

\section{Protocol Description}
\label{sec:protocol} 

This section details the sealed-bid auction protocol, outlining its operational flow, how it achieves the defined security and privacy goals, and analyzing its behavior in adversarial environments.

\subsection{Overview}
\label{sec:overview} 

The protocol enables sealed-bid auctions on public blockchains by separating confidential execution from public settlement. It employs a confidential compute execution layer to handle privacy-sensitive bid processing, verifiable winner determination, and signed-transaction construction - while relying on a public blockchain to enforce settlement.

All value transfers and commitments occur within the native transaction model of the settlement blockchain. The protocol only assumes the integrity of the TEE and the execution guarantees of the confidential compute layer.

Bid confidentiality is preserved until resolution, which is initiated by the enclave through emitting signed settlement transactions. These can be relayed to the settlement blockchain by any participant without coordination. By isolating sensitive logic and minimizing off-chain interaction, the protocol achieves sealed-bid semantics under the trust and infrastructure assumptions defined in Section~\ref{sec:sec-assumptions}, while remaining fully compatible with existing blockchain infrastructure.

\subsection{Auction Lifecycle}

\begin{figure}[t]
  \centering
  \includegraphics[width=0.85\linewidth]{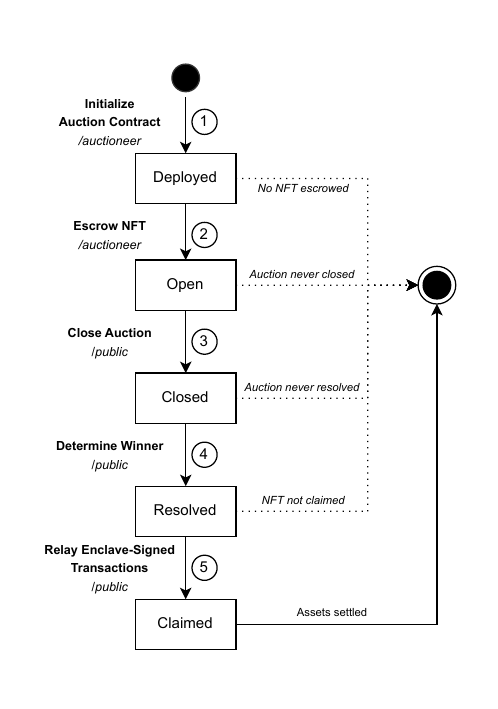}
  \caption{Lifecycle of a sealed-bid auction instance. Each state corresponds to a protocol phase; dotted paths indicate fallback or recovery transitions.}
  \label{fig:lifecycle}
\end{figure}

Figure~\ref{fig:lifecycle} shows the state machine governing each auction instance. The protocol proceeds through five phases: \textsc{Deployed}, \textsc{Open}, \textsc{Closed}, \textsc{Resolved}, and \textsc{Claimed}, each representing a distinct stage in the auction lifecycle.

\paragraph{1.~Deployment (\textsc{Init} $\rightarrow$ \textsc{Deployed}).}
The auctioneer deploys a new auction instance on the execution layer, defining the auction deadline in terms of the settlement-layer block height.

\paragraph{2.~Setup (\textsc{Deployed} $\rightarrow$ \textsc{Open}).}
The auctioneer submits a setup call to the execution layer. The enclave generates a fresh key pair and publicly exposes the corresponding address, which will hold the auctioned asset. The auction transitions to \textsc{Open} once the asset is verifiably escrowed to this address. At that point, prospective bidders can independently confirm that the asset is securely held by the protocol.

\paragraph{~Bidding (\textsc{Open}).}
To place a bid, a participant sends an request to the execution layer, including an encryption key for confidential response delivery. The enclave generates a settlement-layer account, encrypts the address to the bidder’s key, and emits it publicly.  The bidder transfers the bid amount to this address. These payments appear indistinguishable from ordinary on-chain transfers, and the bidder–address mapping remains confidential within the enclave’s sealed state.

\paragraph{3.~Closing (\textsc{Open} $\rightarrow$ \textsc{Closed}).}
Once the auction deadline has passed, any party may initiate a close request. The enclave issues a verifiable off-chain query to the settlement layer interface to confirm that the deadline block has been reached and then finalizes the bidder set.

\paragraph{4.~Resolution (\textsc{Closed} $\rightarrow$ \textsc{Resolved}).}  
Once closed, the enclave resolves the auction by performing verifiable off-chain queries to public settlement-layer interfaces, retrieving balances at the pre-specified deadline block. To enforce the cutoff, transfers confirmed after the deadline block are disregarded when evaluating balances at a given block. Late top-ups may occur but have no effect on the outcome. The enclave constructs and signs the transaction transferring the asset to the winner and others refunding non-winning bidders. It emits the raw transaction payloads for submission by any party. Since each transaction reveals its corresponding funding address, the full bidder set becomes public, enabling verifiability of the result. Resolution requests scale linearly with the number of bidders, as each address must be queried individually. A more efficient resolution strategy with an additional challenge-period is presented in Section~\ref{sec:scalability}.

\paragraph{5.~Settlement (\textsc{Resolved} $\rightarrow$ \textsc{Claimed}).}
Any relayer may submit the enclave-signed transactions to the settlement layer. Recipients are incentivized to ensure inclusion of their own transaction. Relayers cover transaction fees, which may be offset through bundling or sponsorship mechanisms (see Section \ref{sec:implementation}). Delegating publication and fee handling minimizes coordination and keeps the protocol lightweight.

\paragraph{~Finalization and Fallbacks (\textsc{Claimed}).}  
The auction is finalized once all signed settlement transactions have been submitted.  Fallback cases such as missing bids, unclaimed refunds, or an unclaimed asset do not block progress or affect other participants. Such conditions are handled during resolution and do not require additional coordination.

\subsection{Protocol Security Goal Satisfaction}

We revisit the security, usability, and scalability guarantees defined in Section~\ref{sec:goals} and identify the concrete mechanisms through which each is achieved. As summarized in Table~\ref{tab:goal-mechanisms}, these properties emerge from the combination of confidential execution, verifiable attestation, and trustless public settlement. 

\begin{table}[H]
  \centering
  \small
  \renewcommand{\arraystretch}{1.2}
  \setlength{\tabcolsep}{6pt}
  \begin{tabular}{@{}l l@{}}
    \toprule
    \textit{Goal} & \textit{Mechanism} \\
    \midrule
    Confidentiality & Encrypted inputs; sealed enclave state \\
    Binding & One-time escrow address; irreversible L1 settlement \\
    Non-Malleability & Enclave-generated keys; sealed bidder mapping \\
    Enforceability & Signed transactions; permissionless submission \\
    Non-Interactivity & Single enclave call and L1 transfer per bidder \\
    Verifiability & Public attestation; observable state transitions \\
    Scalability & Constant per-bidder cost; linear resolution overhead \\
    \bottomrule
  \end{tabular}
  \caption{Security, usability and scalability goals, and mechanisms by which they are realized.}
  \label{tab:goal-mechanisms}
\end{table}

The guarantees hold under standard conditions, but  deserve clarification. \textbf{Binding} relies on escrow addresses, yet bidders can increase their bid before the cutoff,  as the protocol does not restrict top-ups. \textbf{Enforceability} assumes settlement-layer interfaces behave honestly; if endpoints return false data, resolution may be affected. These issues are addressed in Section~\ref{sec:threats}.

In addition to its security guarantees, the protocol satisfies key functional goals: it is capital-efficient for honest bidders, requires no collateral or penalties, and operates within unmodified blockchain infrastructure. Trust is limited to the attested enclave and the integrity of the confidential compute layer.

\subsection{Threats and Fallbacks}
\label{sec:threats}

\paragraph{Residual TEE Risk.} Despite attestation and sealing guarantees, the protocol inherits the residual vulnerabilities of practical TEE deployments. Architectures such as Intel SGX have been compromised via confidentiality leaks (Foreshadow~\cite{Bulck2018}, \mbox{ÆPIC Leak}~\cite{Borrello2022}), integrity faults (Plundervolt~\cite{Murdock2020}), and attestation subversion (SGAxe~\cite{Schaik2021}). A successful attack may expose internal state including bidder mappings and the private keys of settlement-layer accounts, enabling theft of escrowed assets and full de-anonymization. While hardened runtimes and sealed storage mitigate common side channels and rollback attempts, the enclave remains a single point of failure without cryptographic fallback.

\paragraph{Resolution Integrity via Settlement Layer Interfaces.} To determine settlement layer balances and enforce the auction deadline, the enclave issues verifiable off-chain queries to settlement-layer interfaces, randomly sampling $n$ out of $m$ endpoints from a publicly declared list to avoid reliance on a single provider. Responses are fetched over authenticated channels, emitted for auditability, and checked for consistency. If discrepancies exceed a configurable threshold, the enclave may fallback to querying alternative or trusted sources, like a controlled and centrally operated interface. Although the protocol tolerates partial endpoint misbehavior, a fully colluding subset may misreport balances or block heights, distorting resolution. This risk affects the cutoff enforcement and final balance evaluation. To handle possible asynchrony between the chains, the enclave may also fetch the settlement block header $H$ for auditability or require $\kappa$ confirmations after the deadline block. As discussed in Section~\ref{sec:conclusion}, stronger integrity guarantees may be achieved by hosting interface logic in TEEs.  Misbehavior is detectable and reputationally costly for interface providers, but no cryptoeconomic deterrents are currently enforced. Services like 1RPC\footnote{\url{https://docs.1rpc.io/web3-relay/}} already provide authenticated, privacy-preserving settlement-layer interfaces.

\paragraph{Metadata leakage and bidder linkability.} While bid values remain confidential within the enclave, an observer may attempt to correlate bidder registrations with subsequent on-chain transfers through timing analysis or address patterns. To reduce linkability, bidders are encouraged to randomize the delay between receiving the escrow address and funding it on the settlement layer, and to avoid reusing wallet identities across blockchains. The protocol does not provide full metadata indistinguishability but aims to make linkage non-trivial for adversaries.

\paragraph{Operational fallbacks.}

The protocol guarantees liveness even with passive or malicious participants. If no asset is escrowed, the auction remains non-open; if no bids arrive, the enclave returns the asset to the auctioneer. As balances are fixed at a cutoff block, top-ups after the deadline are ignored. Resolution is economically incentivized: the auctioneer seeks payout, winners seek settlement, and each signed transaction has a beneficiary motivated to broadcast it and cover fees. This ensures that all outcomes eventually materialize without requiring continuous availability or coordination from any actor.

\section{Implementation \& Evaluation}
\label{sec:implementation}  

This section details the protocol’s implementation, formalizes the execution-layer requirements, and evaluates scalability, costs, and design tradeoffs through a working prototype.

\subsection{Execution Layer Requirements}
\label{sec:execution-layer-requirements}

Our sealed-bid auction is independent of any specific confidential compute protocol, but relies on basic properties commonly provided by confidential execution platforms. At its core, it assumes an off-chain computation environment based on a TEE, in which all sensitive auction logic is executed inside an attested enclave.

The execution layer must meet three fundamental requirements. First, it must accept encrypted bidder inputs under an enclave-attested public key. Second, it must maintain sealed, tamper-resistant internal state across the auction lifecycle. Third, it must support verifiable off-chain computation, producing attestable outputs that incorporate both authenticated randomness (e.g., SGX \texttt{RDRAND}) and verifiable queries to settlement-layer interfaces. Several confidential compute blockchains provide dedicated support for these capabilities, as summarized in Table~\ref{tab:exec-layer-capabilities}.

In addition to these runtime guarantees, the execution environment must expose software primitives required to implement the protocol logic: \textbf{keypair generation} for settlement-layer accounts, \textbf{transaction signing} with replay protection, \textbf{RLP encoding} for Ethereum-compatible settlement messages, and \textbf{emits} that deliver outputs confidentially to specific bidders.

\renewcommand{\arraystretch}{1.2}
\rowcolors{2}{gray!2}{white}

\begin{table*}[t]
  \centering
  \small
  \renewcommand{\arraystretch}{1.3}
  \rowcolors{2}{gray!2}{white}
  \begin{tabular}{@{}p{2.6cm}p{2.6cm}p{3.3cm}p{3.3cm}p{3.6cm}@{}}
    \toprule
    \textit{Chain} &
    \textit{Encrypted Inputs} &
    \textit{Confidential State} &
    \textit{Cross-Chain Messaging} &
    \textit{Verifiable Randomness} \\
    \midrule
    SUAVE &
      \ding{51} &
      \ding{51} (\texttt{confidentialStore}) &
      \ding{51} (\texttt{doHTTPRequest}) &
      \ding{51} (\texttt{randomBytes}) \\
    Secret Network &
      \ding{51} &
      \ding{51} &
      \ding{51} (\texttt{secretpath}) &
      \ding{51} (Secret VRF) \\
    Phala Network &
      \ding{51} &
      \ding{51} (Phat Contract) &
      \ding{51} &
      \ding{51} (pRuntime) \\
    Oasis Sapphire &
      \ding{51} &
      \ding{51} &
      \ding{51} &
      \ding{51} (\texttt{randomBytes}) \\
    TEN Protocol &
      \ding{51} &
      \ding{51} &
      \ding{51} (MessageBus) &
      \ding{51} (\texttt{block.prevrandao})\\
    \bottomrule
  \end{tabular}
  \caption{Requirements of confidential compute chains used in our protocol.}
  \label{tab:exec-layer-capabilities}
\end{table*}

\subsection{Prototype Implementation}

We prototype the protocol on Ethereum Sepolia as settlement layer and SUAVE\footnote{\url{https://github.com/flashbots/suave-geth}} as confidential execution layer. SUAVE, now discontinued, was an experimental Flashbots framework extending Ethereum’s stack with TEE-like capabilities. Our prototype builds on the \texttt{Toliman} testnet, a \texttt{go-ethereum}\footnote{\url{https://github.com/ethereum/go-ethereum}} fork with declarative precompiles emulating enclaves, supporting encrypted inputs, sealed state, and authenticated off-chain queries. While TEE functionality is mocked via authenticated APIs, the interface anticipates integration with SGX-backed enclaves.

The protocol uses mainly four SUAVE-specific precompiles:
\begin{itemize}
\setlength\itemsep{0.4em}
\renewcommand\labelitemi{--}
    \item \texttt{confidentialInputs}: processes encrypted input fields in transactions.
    \item \texttt{confidentialStore}: maintains sealed, enclave-protected persistent state.
    \item \texttt{randomBytes}: provides attestable enclave randomness.
    \item \texttt{doHTTPRequest}: performs authenticated off-chain queries.
\end{itemize}

Settlement is triggered by emitting RLP-encoded, enclave-signed Ethereum transactions, which are submitted via Flashbots bundles\footnote{\url{https://github.com/flashbots/ethers-provider-flashbots-bundle}} as transaction-fee sponsoring mechanism. This enables decoupling gas-fee considerations from the enclave's transaction construction, enabling gas-sponsorship and simplifying execution logic. Bidder-specific outputs are encrypted using public keys retrieved via \texttt{eth\_getEncryptionPublicKey} and emitted as ciphertexts. Clients decrypt the responses locally using \texttt{eth\_decrypt}\footnote{\url{https://docs.metamask.io/wallet/reference/json-rpc-methods/eth_decrypt/}}, which, despite being deprecated, remains widely supported in Ethereum wallet environments and suffices for prototyping confidential output delivery.

The complete implementation is available at \url{https://github.com/TimmMu/tee-confidential-auctions}.

\subsection{Gas Cost Analysis and Scalability}
\label{sec:scalability}

We evaluate gas consumption on both the confidential execution layer (SUAVE) and the settlement layer (Sepolia). All contracts were compiled with \texttt{solc} version~0.8.20 using \texttt{10,000} optimizer runs. Gas measurements on SUAVE’s testnet are indicative, as custom precompiles such as \texttt{doHTTPRequest} are currently priced at \texttt{1,000} gas units, underestimating realistic overhead. To address this, we additionally report a adjusted scenario in which each off-chain HTTP request incurs a conservative cost of \texttt{100,000} gas.

\begin{table}[H]
  \centering
  \label{tab:gas-cost-breakdown-4}
  \begin{tabular}{|l|r|r|}
    \hline
    \textbf{Operation} & \textbf{Ethereum (L1)} & \textbf{SUAVE} \\
    \hline
    Deploy auction         & 0       & 3,849,426 \\
    Start auction          & 70,618  &   124,324 \\
    Submit bid (avg)       & 21,000  &   271,160 \\
    End auction            & 0       &   804,800 \\
    Claim valuable (avg)   & 21,000  &        0 \\
    \hline
  \end{tabular}
  \caption{Gas cost per operation in the 4-bidder auction experiment.}
\end{table}

\begin{figure}[h]
    \centering
    \includegraphics[width=0.95\linewidth]{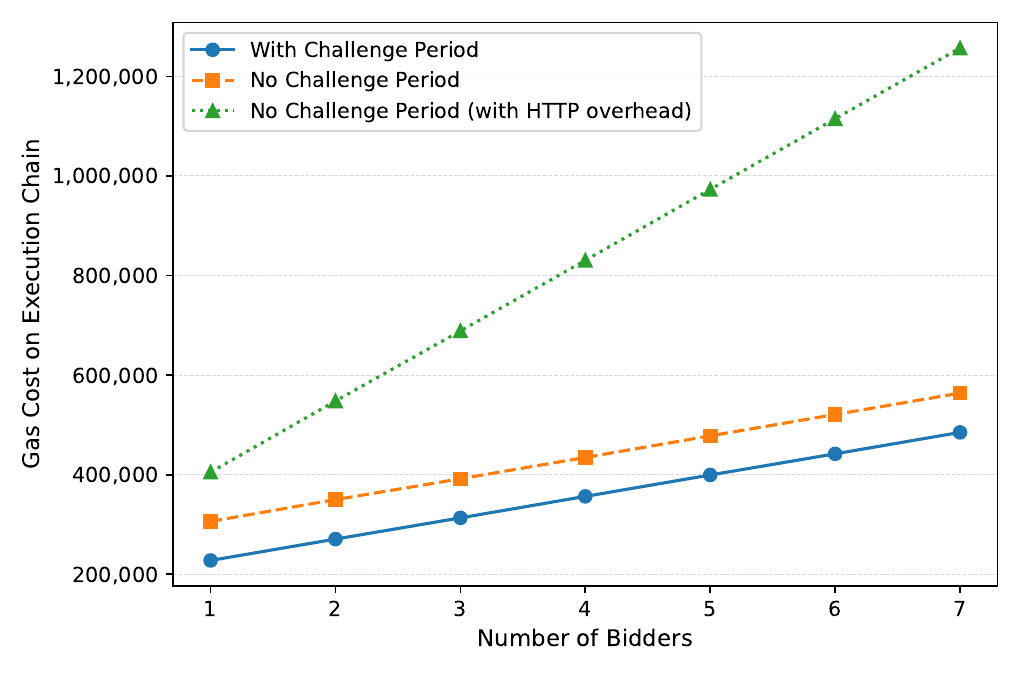}
    \caption{Gas consumption on the SUAVE blockchain as a function of the number of bidders in a sealed-bid auction.}
    \label{fig:gas-cost-plot}
\end{figure}

Table~\ref{tab:gas-cost-breakdown-5} reports the per-operation gas costs for a four-bidder auction. On the settlement layer, costs are minimal and are limited to standard asset transfers. On the execution layer, bidders incur costs for obtaining escrow addresses. The end-auction is initiated by a single party, who covers its full gas cost. As illustrated in Figure~\ref{fig:gas-cost-plot}, this call scales linearly with the number of bidders, since each balance must be verified via an off-chain query at a fixed block height.

The linear growth in off-chain balance queries with the number of bidders poses a core scalability limitation. To address this, we implement a \emph{proposer-based resolution mechanism}. After the auction deadline, the contract enters a \emph{proposal phase} in which any party can suggest a candidate winner. For each proposal, the enclave verifies two conditions: (i) the bid exceeds the current highest, and (ii) the bidder’s escrow address held sufficient funds at the cutoff block. This reduces the number of off-chain queries to one per proposal. However, this comes at a cost: \emph{funds remain temporarily locked}, and \emph{non-interactivity is lost}, as bidders must observe the challenge phase and may need to contest invalid proposals.

Figure~\ref{fig:gas-cost-plot} shows gas usage for both resolution variants, including the scenario with increased precompile costs. A detailed breakdown for the proposer-based implementation is provided in Table~\ref{tab:gas-cost-breakdown-5} in the appendix.

\section{Related Work}
\label{sec:relatedwork}  

A wide range of proposals aim to realize sealed-bid auctions on blockchain infrastructure. We distinguish three main design mechanisms: \textbf{(i)}~\emph{commit–reveal protocols}, \textbf{(ii)}~\emph{cryptographic approaches with semi-trusted parties or MPC}, and \textbf{(iii)}~\emph{trusted hardware–based solutions}.

\paragraph{Commit–reveal protocols.}
These schemes~\cite{Krol2023, Braghin2020} split the bidding process into a commitment and a subsequent reveal phase. While simple and gas-efficient, they suffer from liveness issues: any bidder can stall the auction by withholding their opening. To mitigate this, some designs introduce deposits, force-reveal mechanisms, or slashing penalties, increasing protocol complexity and often requiring off-chain enforcement or social assumptions.

\paragraph{Cryptographic constructions with semi-trusted actors.}
A range of proposals replaces interactive bidding with cryptographic proofs, while often relying on designated parties or tolerating substantial verification costs.
\emph{Galal \& Youssef}~\cite{Galal2018} commit bidders via Pedersen commitments and require them to encrypt openings to a designated auctioneer, who reveals the winner using Chaum–Pedersen interval proofs. Although deposits punish misbehavior, the auctioneer learns all bids and on-chain verification incurs significant gas overhead.
\emph{Strain}~\cite{Blass2017} uses Goldwasser–Micali encryptions and pairwise secure comparisons between bidders, with a semi-honest contract owner verifying zero-knowledge proofs and publishing the ranking. While bid values remain secret, the full order becomes public, scalability is limited due to quadratic proof complexity, and the approach lacks on-chain settlement enforcement.
\emph{Riggs}~\cite{Tyagi2023} achieves a trust-free Vickrey auction using Bulletproofs and RSA-based timed commitments that allow force-opening. Bids remain private before closure, but each bid costs $\approx$2.5M gas and the maximum bid ceiling is publicly exposed.
\emph{ZeroAuction}~\cite{Zhang2024} defers execution of encrypted transactions until a consensus-level timeout, achieving perfect bid privacy until expiry without deposits or interaction. However, it requires modifications at the consensus layer enabling the protocol.

\paragraph{Trusted hardware–based solutions.}
Another class of protocols leverages TEEs to offload private logic into isolated enclaves.
\emph{Town Crier}~\cite{Zhang2016} pioneered the use of SGX enclaves in the blockchain context as authenticated relays between smart contracts and HTTPS data sources. By verifying attested enclave code and encrypting requests, it enabled data integrity without trusting the operator.
\emph{Trustee}~\cite{Galal2019} follows this relay-design by executing auction logic inside an SGX enclave, determines the winner, and submits a signed transaction to the blockchain. While gas-efficient, the protocol relies on a centralized relay, which may censor or delay resolution. Our protocol eliminates this dependency by anchoring settlement directly on a confidential compute blockchain.

\section{Conclusion}
\label{sec:conclusion}

We presented a sealed-bid auction protocol that achieves confidentiality, enforceability, and auditability without trusted intermediaries. By combining TEE-backed execution with on-chain settlement, the protocol preserves sealed-bid semantics while remaining compatible with existing infrastructure. Our implementation on SUAVE and Ethereum demonstrates practical feasibility with minimal on-chain overhead. Although SUAVE has been discontinued, the protocol remains agnostic to the underlying infrastructure and deployable on any confidential compute blockchain meeting basic execution and verification requirements.\newline
Resolution still depends on settlement-layer interfaces, introducing residual trust; enclave execution of interface logic could mitigate this. Integrity is probabilistic but reinforced by reputational disincentives. As with all TEE-based systems, enclave compromise and side-channel leakage remain open challenges.\newline
Overall, our results show that sealed-bid auctions with strong privacy and verifiable settlement are feasible today, without protocol changes, multi-phase interaction, or centralized trust, enabling confidential price discovery for DeFi use cases such as NFT auctions and tokenized real-world assets like carbon credits.

\begin{acks}
We thank Lennard Laurenz Harte, Doganay Mehmet Harun and Yelkanat Irem Ecem for their work on prototype implementation. We also thank the anonymous reviewers for their feedback.
\end{acks}

\bibliographystyle{ACM-Reference-Format}
\bibliography{defi25-2}

\appendix
\newpage

\section{Appendix}
\subsection{Gas Cost Breakdown}

\begin{table}[H]
  \centering
  \begin{tabular}{|l|r|r|}
    \hline
    \textbf{Operation} & \textbf{Ethereum (L1)} & \textbf{SUAVE Blockchain} \\
    \hline
    Deploy auction           & 0       & 4,122,288 \\
    Start auction            & 70,618  &   55,403 \\
    Submit bid (avg)        & 21,000  &   271,998 \\
    End auction              & 0       &   398,827 \\
    Register winner (avg)   & 0       &   154,283 \\
    Claim valuable (avg)    & 21,000  &        0 \\
    \hline
  \end{tabular}
  \caption{Gas cost per operation in the 4-bidder proposed-based experiment. Bidding and resolution involve SUAVE Toliman and Ethereum Sepolia testnet.}
  \label{tab:gas-cost-breakdown-5}
\end{table}

\subsection{Comparison with Related Sealed-Bid Auction Protocols}

\begin{table*}[h]
  \centering
  \small
  \renewcommand{\arraystretch}{1.4}
  \begin{tabular}{
    p{2.8cm}
    p{1.9cm}
    p{2.2cm}
    p{1.9cm}
    p{2.1cm}
    p{2.4cm}
    >{\columncolor{lightgray}}p{2cm}
  }
    \toprule
    \textbf{} &
    \textbf{Strain~\cite{Blass2017}} &
    \textbf{G\&Y~’18~\cite{Galal2018}} &
    \textbf{Trustee~\cite{Galal2019}} &
    \textbf{Riggs~\cite{Tyagi2023}} &
    \textbf{ZeroAuction~\cite{Zhang2024}} &
    \textbf{\textbf{Our protocol}} \\
    \midrule
    \makecell[l]{Confidentiality} &
      \makecell[l]{leaks order} &
      \makecell[l]{auctioneer sees all} &
      \makecell[l]{\checkmark} &
      \makecell[l]{values hidden, \\ cap leaks} &
      \makecell[l]{\checkmark} &
      \makecell[l]{\checkmark} \\

    Non-Interactivity. &
      \ding{55} &
      \ding{55} &
      2 tx &
      \checkmark &
      \checkmark &
      \checkmark \\

    Binding &
      \makecell[l]{\checkmark} &
      \makecell[l]{\checkmark} &
      \makecell[l]{\checkmark} &
      \makecell[l]{\checkmark} &
      \makecell[l]{weak} &
      \makecell[l]{\checkmark} \\

    Enforceability \newline (Enforceable reveal) &
      quorum shares &
      deposits &
      TEE &
      force-open &
      built-in &
      TEE \\

    Enforceability \newline (Payment enforcement) &
      \ding{55} &
      deposits &
      contract &
      collateral &
      atomic in tx &
      escrow accounts \\

    Trusted element &
      judge &
      auctioneer &
      TEE + relay &
      RSA setup &
      decrypt committee &
      TEE + conf. compute blockchain\\

    On-chain gas / bid &
      high &
      high &
      very low &
      very high &
      very low &
      very low \\

    Deployable today &
      \checkmark &
      \checkmark &
      \makecell[l]{\checkmark} &
      \checkmark &
      \makecell[l]{\ding{55}} &
      \checkmark \\
    \bottomrule
  \end{tabular}
  \caption{Comparison of sealed-bid auction protocols.}
\end{table*}

\end{document}